# OPTO-ACOUSTIC OSCILLATOR USING SILICON MEMS OPTICAL MODULATOR

*Suresh Sridaran and Sunil A. Bhave*
OxideMEMS Lab, Cornell University, Ithaca, NY, USA

## ABSTRACT
We show operation of a silicon MEMS based narrow-band optical modulator with large modulation depth by improving the electro-mechanical transducer. We demonstrate an application of the narrowband optical modulator as both the filter and optical modulator in an opto-electronic oscillator loop to obtain a 236.22 MHz Opto-Acoustic Oscillator (OAO) with phase noise of -68 dBc/Hz at 1 kHz offset.

## INTRODUCTION
Hybrid oscillators such as the optical delay line based Opto-Electronic Oscillators (OEOs) have superior phase noise compared to quartz and acoustic MEMS oscillators in the 1-30 GHz frequency range [1]. The OEO uses an RF filter and an electro-optic modulator for filtering the RF signal and up-converting it to optical frequencies. The RF signal is carried on the optical signal using a long optical fiber as the delay line. The RF signal is then reconverted into the electrical domain using an optical detector and fed back as input for the filter as shown in figure 1. Multiple RF frequencies are supported in the combined optical/electrical loop out of which one frequency is chosen for oscillation by the RF filter. The superior phase noise is due to the long energy storage times provided by the optical fiber. The equivalent quality factor of the delay line is given by the expression [2] $Q_d = 2\pi \cdot f_{osc} \tau_d$ where $f_{osc}$ is the RF oscillation frequency and $\tau_d$ is the delay introduced by the delay line. For the case of $f_{osc}$ of 10GHz and a length of fiber spool of 2 km corresponding to a delay $\tau_d = 10\mu s$, a $Q_d$ of 62,800 is obtained which is 20x larger [3] than the quality factor of resonators obtainable at that frequency.

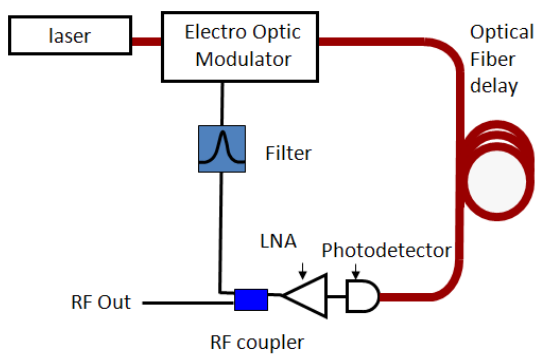

**Figure 1**: Schematic of an Opto Electronic Oscillator (OEO)

Efforts to miniaturize the OEO have focused on replacing the long optical fiber delay with an optical whispering-gallery mode high-Q resonator [4, 5] or a fiber ring resonator [6]. Previous efforts at using the high-Q resonator involve machined disk resonators made from lithium niobate [4], fused silica [5], MgF$_2$ [5] or CaF$_2$ [4] with optical quality factors ($Q_{opt}$) of $10^8$ for modes around 1550nm. These resonator-based OEOs utilize the resonator's optical free spectral range in order to set the oscillation frequency and work typically around 10GHz. The equivalent delay-quality factor at the oscillation frequency for these structures is given by [5] $Q_d = \dfrac{f_{osc} \cdot Q_{opt}}{f_{opt}}$ which works out to a $Q_d$ of 5,180 for oscillation at 10GHz for a $Q_{opt}$ of $10^8$. These delay-quality factors are similar to the $Q_{mech}$ limit of Silicon which is around 10,000 from 1GHz to 10GHz [7]. Therefore, using a MEMS filter in the opto-electronic oscillator loop would result in oscillations with an effective quality factor significantly better than that achievable with a filter in just a single domain.

In addition to the delay, the optical modulator that needs to be scaled down is either integrated along with the high-Q resonator used for the delay [4] or implemented using chip-scale silicon photonic modulators [8]. In this work, we demonstrate a silicon MEMS based narrowband optical modulator employed as both the MEMS filter and optical modulator in an opto-electronic oscillator loop to obtain a 236.22MHz Opto-Acoustic Oscillator (OAO). The hybrid acousto-optic filter and modulator reduces the inefficiencies of repeated domain conversion and is a key enabler for chip-scale miniaturization of the opto-electronic oscillator that has hitherto been limited by the lack of an efficient modulation and filtering mechanism.

## RF MEMS BASED OPTICAL MODULATOR
### Operating Principle
The MEMS based narrowband optical modulator consists of two mechanically coupled disk resonators as shown in figure 2. The dilatational motion in the MEMS resonator (on the left) is driven by 150nm air gap capacitive actuators surrounding the disk. The vibrations are coupled through a coupling beam [9] to a second disk (on the right) which acts as the optomechanical resonator.

When continuous wave laser light enters a waveguide next to the disk with an optical wavelength close to the optical resonance of the disk, it evanescently couples into the disk and builds up in intensity within the disk. Turning on the electrical actuation produces radial vibrations that introduce an optical phase shift thereby changing the wavelength of optical resonance of the disk. A radial displacement $\Delta r$ results in a shift in optical resonance wavelength equal to $\Delta\lambda = (\Delta r/R)\lambda_o$ where R is the radius of the disk and $\lambda_o$ is free space wavelength. The shifting of the resonance wavelength causes a change in the light intensity inside the disk which manifests itself as modulation of the light at the output waveguide. The frequency of this intensity modulation is the same as that of the radial vibrations of the disk. The radial vibrations of the disk are, in turn, excited only when the mechanical disk resonator is electrically actuated at the mechanical resonance frequency. Thus, the optical modulation has a narrowband response to the input electrical frequency, with the bandwidth specified by the quality factor of the mechanical resonator.

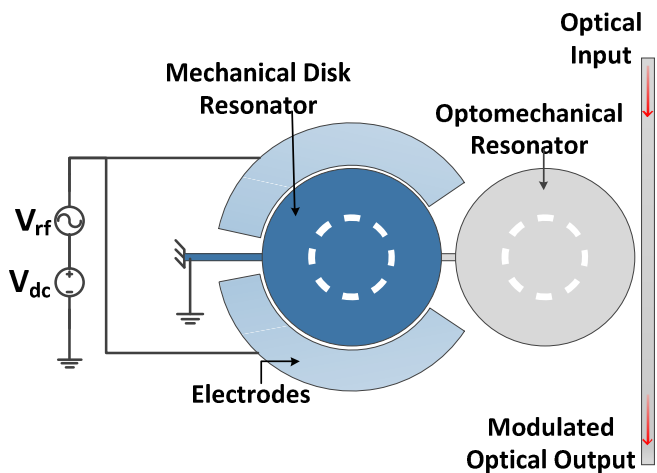

*Figure 2*: Schematic of the MEMS based optical modulator with mechanical resonator actuated by air gap capacitors on the left coupled to an optomechanical resonator on the right through a coupling beam.

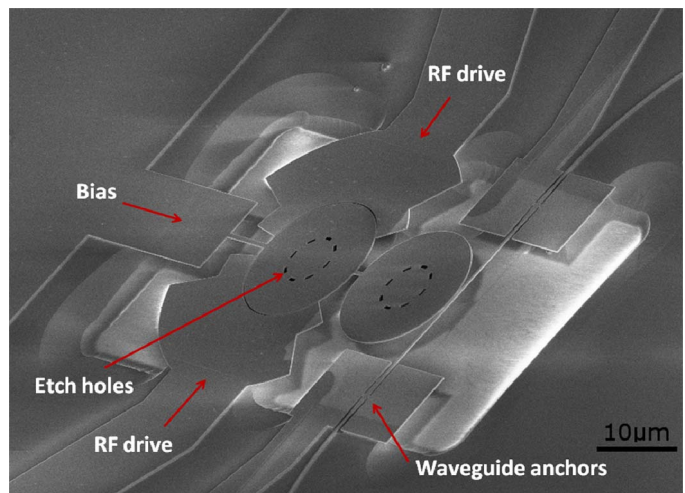

*Figure 3*. SEM of optical modulator fabricated on 220 nm thick Si on 3 μm thick buried oxide SOI wafer. The electrodes are spaced 150nm away from the 10 μm radius resonators, which have etch holes to allow buffered HF to undercut them. The 350 nm wide waveguide is suspended 150 nm away from the optical disk resonator using anchors designed to reduce the scattering loss.

**Fabrication and Characterization**

The device is fabricated using a three mask process on silicon-on-insulator wafers with 250nm device layer and 3μm buried oxide. The first mask defined using electron beam lithography is used to pattern the device layer, while the second mask is used to selectively ion implant the mechanical resonator and electrical routing beams to reduce their electrical resistance. The third mask is used to specify release regions for the disk. The structures are released by a timed etch in buffered oxide etchant and dried using critical point dryer to prevent stiction. A scanning electron microscope image of a fabricated device is shown in figure 3.

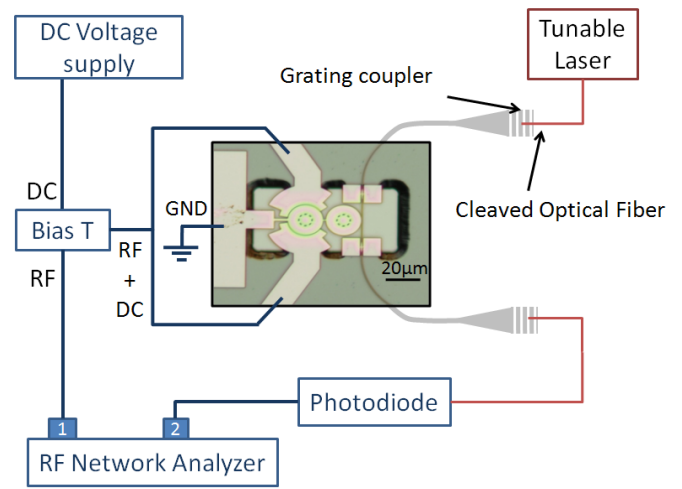

*Figure 4*.Setup for measuring the transmission response of the silicon RF MEMS based optical modulator.

Light from a tunable laser polarized parallel to the substrate is coupled to the waveguides using gratings to measure the optical transmission characteristics of the disk resonators. The loss at each of the gratings is 10.8dB. Anchor structures are used to hold the waveguide in place during the release process and reduce the optical scattering loss [10]. An optical resonance with a quality factor of 30,200 at a wavelength of 1550.9nm is measured.

A setup as shown in figure 4 is used to observe the characteristics of the optical modulator. The optical input wavelength is biased at the half maximum point of the disk optical resonance and the optical signal is converted into the electrical domain using a high speed photodiode with a gain of 4500V/W. The transmission spectrum corresponding to the electro-optical conversion gain is measured on a RF network analyzer and is shown in figure 5. The transmission is highest at the mechanical resonance frequency of the device at 236.22MHz. The response shows the presence of two modes of vibration due to splitting of the radial vibration mode due to the coupling beam. The two modes which correspond to in-phase and out-of-phase motion of the disks are shown in the insets of figure 5. Increasing the DC bias voltage results in a larger actuation force and thereby an increase in amplitude of vibration. This leads to larger modulation of optical power for a given input RF power and is seen as a higher value of the transmission (S21). This confirms that the modulation is indeed caused by electrostatic actuation of the mechanical resonator disk. The quality factor for the mechanical resonance at 236.22MHz is 900 measured in air with a modulated optical power of -25.64dBm for a 5dBm RF input and a 3dBm optical input. As the electrical routing beams and resonators are all designed in the same layer, the disk is electrically grounded by the beam that is connected to its edge. The drawback to this scheme is reduced $Q_{mech}$ resulting from placing the beam at the maximum displacement point of the radial mode of the disk thereby increasing the anchor loss.

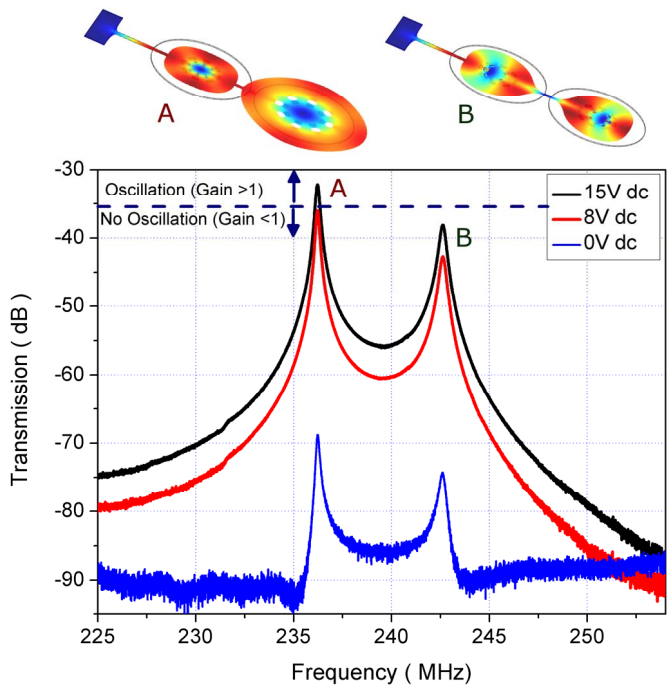

*Figure 5. Transmission of modulator measured on the network analyzer for RF input power of 5dBm, and 3dBm CW laser power with DC bias of 0V (blue), 8V (red) and 15V(black). Any point on the transmission spectrum above the dashed line satisfies gain ≥ 1 for oscillation while resonances below the line are suppressed. Insets A and B show the displacement mode shape obtained for resonances at 241.3MHz (236.22MHz measured) and 248.3 MHz (242.65MHz measured) respectively using COMSOL.*

The power measured at the network analyzer is dependent on the average optical power incident on the photodetector. Due to losses in the grating, the average optical power incident on the detector is 20dB less than the 3dBm input level. This loss shows up as a reduced RF power level at the output of the photodetector, thus necessitating a high gain detector. A better figure of merit for the modulator compared to the transmission at the RF network analyzer is the extinction ratio. The extinction ratio is defined as the amplitude of optical power modulation around the average optical power at the bias point and is given by $M = 10\log_{10}(P_{max}/P_{min})$. The extinction ratio for the MEMS based modulator is 6dB at a DC bias of 20V and RF power of 5dbm. If the grating losses are ignored and an average optical power of 10mW passing through the modulator is assumed to be incident on a unity gain detector, it would correspond to a RF power at the output of the detector of 1dBm.

## OPTO-ACOUSTIC OSCILLATOR (OAO)

An opto-acoustic oscillator is realized using the MEMS based optical modulator by using the device in a closed loop as shown in figure 6. In order to achieve oscillations, it is necessary to achieve unity open loop gain with a phase shift of an integral multiple of 2π. In order to achieve the unity gain criterion, a low noise amplifier with a gain of 36dB is used to amplify the signal out of the detector. To meet the phase condition, this signal is then routed through a phase shifter before being fed back into the MEMS-based optical modulator.

The existence of multiple mechanical vibration modes can be seen in modulator transmission response in figure 5. In order to select and stabilize the targeted oscillation at the 236.22 MHz, we choose the amplifiergain and DC bias of the transducer such that the oscillation conditions are satisfied only at that frequency and at none of the other frequencies. For bias voltages above 8V, the modulator transmission shown in figure 5 lie above the dashed line representing -36dBm. For these bias voltages, the open loop gain is greater than 1 and oscillations can be sustained. Figure 7 shows the RF output of the oscillator observed on the spectrum analyzer for different bias voltages. As expected, the oscillation threshold is at a bias voltage of 8.2V with oscillations sustained above that bias voltage. The phase noise of the oscillator is measured with a signal source analyzer. For a 15V DC bias, an output RF power of 6.5dBm is obtained at 236.22MHz and the measured phase noise is -68 dBc at 1 kHz offset from carrier frequency, as shown in figure 8. The oscillation waveform characterized using an oscilloscope is shown in figure 9 with a time period of 4.23ns.

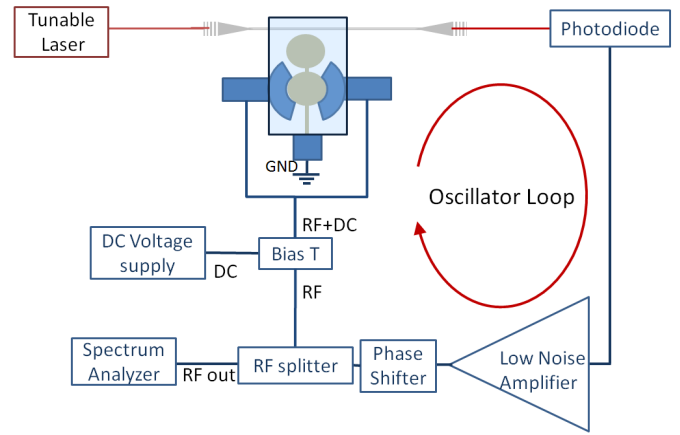

*Figure 6. Schematic of the Opto-Acoustic Oscillator setup with commercial off-the-shelf RF components.*

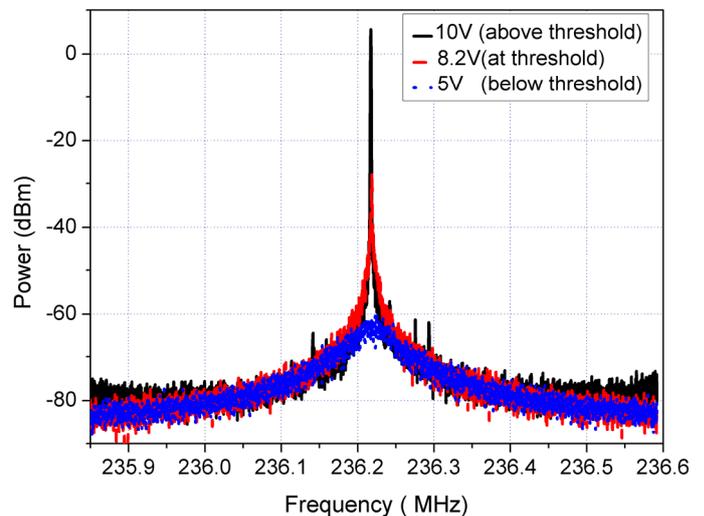

*Figure 7. RF output of the oscillator as measured on the spectrum analyzer with a power output of 4.8dBm at 236.22MHz with a DC bias of 10V.*

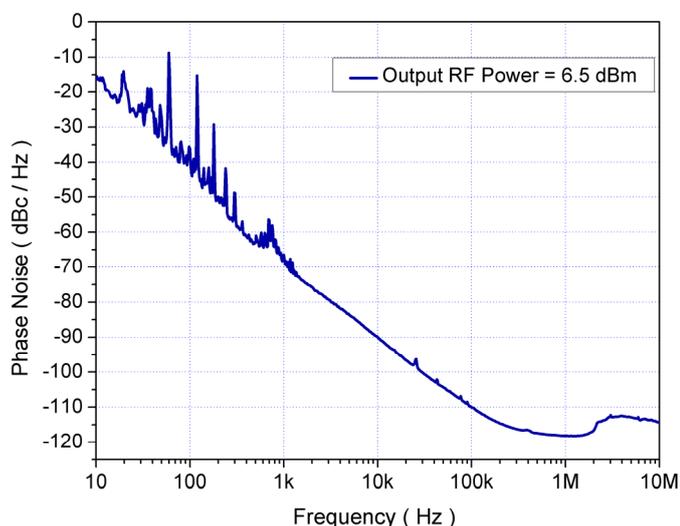

*Figure 8. Phase Noise of the oscillator measured using Agilent Signal Source Analyzer at 236.22MHz with optical input power of 2mW and DC bias of 15V*

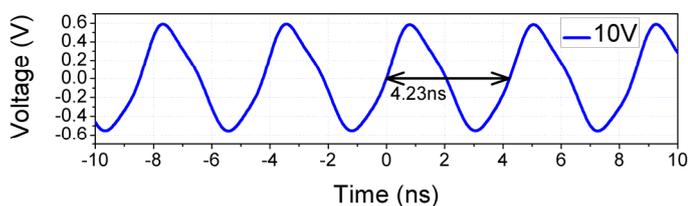

*Figure 9. OAO output measured on an oscilloscope with a DC bias of 10V.*

## CONCLUSION

We have demonstrated a MEMS based narrowband silicon optical modulator operating at a frequency of 236.22MHz with an extinction ratio of 6dB. There is no DC power dissipated in this device as the DC bias is applied across a capacitor. As the optical response to the electrical input is shaped by the mechanical transfer function, the device is similar to a RF filter in series with an optical modulator. Using this hybrid filter and modulator, we have demonstrated an OAO operating at 236.22 MHz with 6.5dBm output RF power for a DC bias of 15V and an input optical power of 3dBm. The phase noise of this oscillator at 1kHz offset from the carrier frequency is -68dBc/Hz.

## ACKNOWLEDGEMENTS
The authors wish to thank the Semiconductor Optoelectronics Group and the Cornell Nanophotonics Group. This work was supported by the DARPA Young Faculty Award, DARPA/MTO's ORCHID program and Intel and was carried out in part at the Cornell NanoScale Science and Technology Facility.

## CONTACT
*Suresh Sridaran, ss625@cornell.edu